\documentclass[twocolumn]{aastex701}
\usepackage[T2A]{fontenc}

\begin{document}

\title{Dynamical response of a protoplanetary disk to an FU Ori-type outburst}

\author[orcid=0000-0002-1176-9048, sname='Roman Zolotarev']{Roman Zolotarev}
\affiliation{Institute of Astronomy, Russian Academy of Sciences, Pyatnitskaya Str. 48, Moscow 119017, Russia}
\affiliation{Main Astronomical Observatory at Pulkovo, Pulkovskoe Sh. 65/1, St. Petersburg 196140, Russia}
\email[show]{rzolotarev@inasan.ru}

\author[orcid=0000-0002-8150-2795, sname='Yaroslav Pavlyuchenkov']{Yaroslav Pavlyuchenkov}
\affiliation{Institute of Astronomy, Russian Academy of Sciences, Pyatnitskaya Str. 48, Moscow 119017, Russia}
\affiliation{Main Astronomical Observatory at Pulkovo, Pulkovskoe Sh. 65/1, St. Petersburg 196140, Russia}
\email{pavyar@inasan.ru}

\author[orcid=0000-0002-4324-3809, sname='Vitaly Akimkin']{Vitaly Akimkin}
\affiliation{Institute of Astronomy, Russian Academy of Sciences, Pyatnitskaya Str. 48, Moscow 119017, Russia}
\affiliation{Main Astronomical Observatory at Pulkovo, Pulkovskoe Sh. 65/1, St. Petersburg 196140, Russia}
\email{akimkin@inasan.ru}

\begin{abstract}
Episodic accretion events onto young stars can significantly change the global structure of their surrounding protoplanetary disks. Using radiation hydrodynamical simulations with the HURAKAN code, we study the response of the outer passive disk to a powerful FU Ori-type outburst. Disk heating initiates a factor-of-two increase in the vertical scale-height, which develops on a local keplerian timescale. This effectively launches an outward travelling wave with a bump whose radial position changes with time as $r_{\rm b}\sim t^{2/3}$. After the outburst ends, new waves are launched as the disk structure re-equilibrates toward a quiescent state. The vertical expansion of the disk introduces significant time-variable extinction of the central source emission. Disks with inclinations in the range $40^{\circ}<i<60^{\circ}$ go from weakly ($A_{\rm V}<1$) to a strongly embedded state ($A_{\rm V}\sim10^2$), with the extinction decaying on a $\sim10^2$~yr timescale after the outburst. A fingerprint of the outburst-launched shadowing wave is a long-lived drop or a transient dip in the luminosity curve occurring several years after the outburst begins. This apparent luminosity variation is relatively large, up to $1^m-3^m$ in the V band, and is not related to the intrinsic accretion variability.

\end{abstract}


\section{Introduction}

The formation of stars is inherently linked to the evolution of their circumstellar disks. Although the stellar embryo (the second hydrostatic core) and its circumstellar disk form during a relatively brief phase of protostellar cloud collapse, the subsequent accretion from the disk onto the protostar takes an order of magnitude longer. Both observational and theoretical studies increasingly show that accretion from the circumstellar disk onto the young star is highly variable~\citep{2023ASPC..534..355F}. Such episodic accretion results in strong luminosity variations in the central source, which consists of the star and the inner active part of the disk. The most extreme variable sources associated with non-stationary accretion are FU Ori-type (FUors) and EX Lup-type stars (EXors). FUors reach luminosities of several hundred solar luminosities over a period of up to several years and are characterized by an outburst phase lasting from tens to, possibly, hundreds of years, whereas EXors reach peak luminosities of tens of solar luminosities with bursts lasting for several months~\citep{1996ARA&A..34..207H, 2014prpl.conf..387A, 2023ASPC..534..355F}.

Different physical mechanisms are considered as causes of non-stationary accretion: 1)~the formation of clumps in a gravitationally unstable disk, followed by their migration and disruption by the central star~\citep{2015ApJ...805..115V}; 2)~the development of various physical instabilities, such as thermal~\citep{1994ApJ...427..987B,2024MNRAS.530.1749N} and magnetorotational~\citep{2009ApJ...701..620Z,2014ApJ...795...61B} instabilities, providing enhanced viscosity in the inner disk; 3)~perturbation by external sources (accretion from filaments, flybys)~\citep{2019A&A...628A..20D,2025AstL...51..310G}. The selection of the proper mechanism should be made by comparing model predictions with observational data. Great progress has been made in this respect, e.g.~\citet{2021A&A...647A..44V,2024MNRAS.528.2182N,2025ApJ...995..129S}. However, many details of the involved accretion-inducing mechanisms remain to be explored.

The energy of FU Ori-type events greatly exceeds the thermal energy of a protoplanetary disk prior to the outburst. The sudden change in accretion luminosity is thought to originate from the inner, sub-au active disk~\citep{2021A&A...646A.102L, 2022A&A...663A..86L,2026A&A...707A.296C}, but it should significantly affect the global disk structure in several ways. The primary response of the outer disk is thermal restructuring, which occurs on a timescale of months to years~\citep{2024A&A...688A...8W,2025A&A...700L..24L}. The rise in disk temperature initiates active chemical evolution, with the most notable effect being the evaporation of volatile species from dust mantles~\citep{2017A&A...604A..15R,2018ApJ...866...46M, 2018A&A...617L...7S, 2019NatAs...3..314L, 2019MNRAS.485.1843W, 2021A&A...646A...3L, 2022ARep...66..393B}.

The dynamics of the outer disk during FU Ori-type outbursts are much less studied. \cite{2018A&A...617L...7S} used radiation-hydrodynamics (RHD) simulations to show that circular surface waves emerge during outbursts and may show up in scattered light images as bright and dark concentric rings. In our previous work \citep{2025A&A...700L..24L}, we used the radiative module of the HURAKAN RHD code to study time-dependent heating and cooling of a protoplanetary disk during FU Ori events, assuming a non-evolving disk density distribution. In this paper, we employ both the time-dependent RT and hydrodynamical modules of the HURAKAN code to study the disk temperature and density evolution self-consistently, and how this may be reflected in observed continuum emission.

\section{Model} \label{sec:mod}

Our simulations consist of two stages. At the first stage, we calculate steady state protoplanetary disk structure assuming thermal equilibrium and hydrostatic equilibrium in the vertical direction. At the second stage, we start to vary the luminosity of the central source in accordance with a given outburst profile and calculate the thermal and hydrodynamical evolution of the disk. The parameters of the protoplanetary disk, the nascent star, and the outburst luminosity profile are related to the FU Ori object itself and are mostly adopted from the paper by~\cite{2025A&A...700L..24L}. We briefly summarize them below.

\begin{figure}[ht!]
\begin{center}
\includegraphics[width=0.99\linewidth]{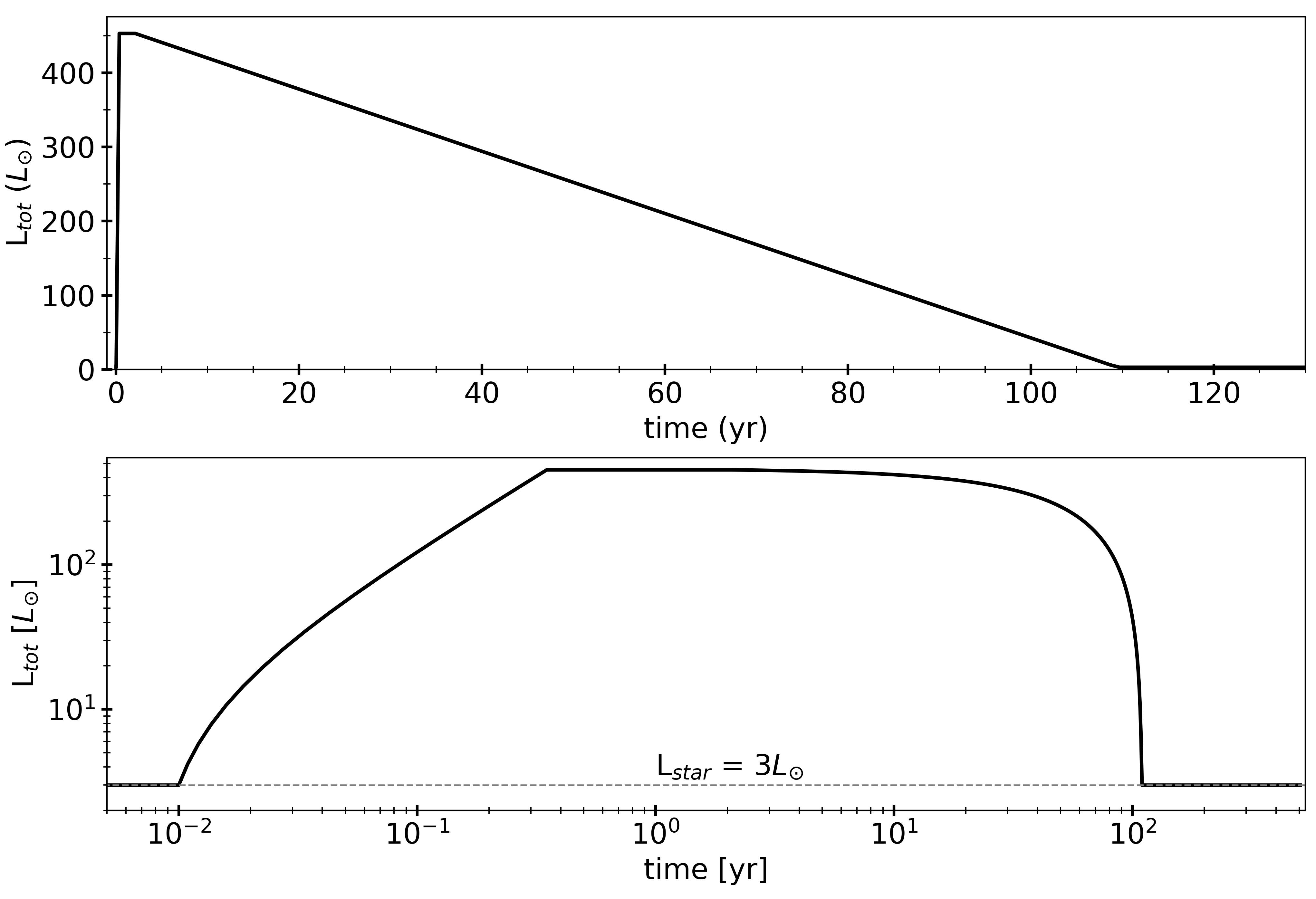}
\end{center}
\caption{Postulated luminosity curve of the central source.}
\label{fig-curve}
\end{figure}

The stellar parameters are $M_{*}=0.6 M_{\odot}$, $R_{*} =3.6R_{\odot}$ and $T_{*}=4000$~K. The central source luminosity is the sum of two black-body components representing the intrinsic stellar and accretion-induced radiation. The stellar component is not evolving ($L_{*}=3\,L_{\odot}$), while the accretion source with effective temperature $T_{\rm acc} = 10^4$~K is variable according to a prescribed outburst luminosity curve. The adopted accretion luminosity profile is based on linear fits of FU Ori luminosity curve observations and consists of a sharp 4-month increase from the pre-outburst value of $0.05\,L_{\odot}$ to a peak value of $450\,L_{\odot}$, which then linearly decreases back to the pre-outburst value over 110~years \citep{2025A&A...700L..24L}, see Fig.~\ref{fig-curve}.

We assume that the active disk regions responsible for the accretion luminosity outburst are confined to the inner sub-au region, and we do not include them into our computational domain. Thus the modeled disk is assumed to be passive, i.e. it does not have significant internal heating related to viscous dissipation. In our model, the expected inner active regions affect the outer disk only by radiation via postulated light curve of the central source. This approach corresponds to ``model~P'' introduced in~\citet{2025A&A...700L..24L}.

The initial surface density distribution in the disk is given by:
\begin{eqnarray*}
    \label{eq:Sigma}
    &\Sigma (r) = \Sigma_0 \left(\frac{r}{1\,\rm{au}}\right)^{-\gamma} \exp \left[-\left(\frac{r}{r_{\rm out}}\right)^{2-\gamma} \right]\times\\
    &\exp \left[-\left(\frac{r}{r_{\rm in}}\right)^{\gamma-2} \right],
\end{eqnarray*}
where $\Sigma_0=163$~g\,cm$^{-2}$, $\gamma=1$, $r_{\text{in}} = 2$~au, $r_{\text{out}} = 100$~au. The corresponding disk mass is $0.01 M_{\odot}$. The dust-to-gas mass ratio is assumed to be uniform throughout the disk and equal to 0.01. The initial vertical structure of the disk is found via successive iterations between calculations of thermal and hydrostatic equilibrium. Initially, the disk region is supposed to be in Keplerian rotation with $v_{\phi}(R,\theta)=\sqrt{GM_*/R}\sin \theta$, where $R$ and $\theta$ are the spherical coordinates of the element. We assume a floor value of $10^4$~cm$^{-3}$ for the gas number density in the initial conditions.

\begin{figure}[ht!]
\begin{center}
\includegraphics[width=0.99\linewidth]{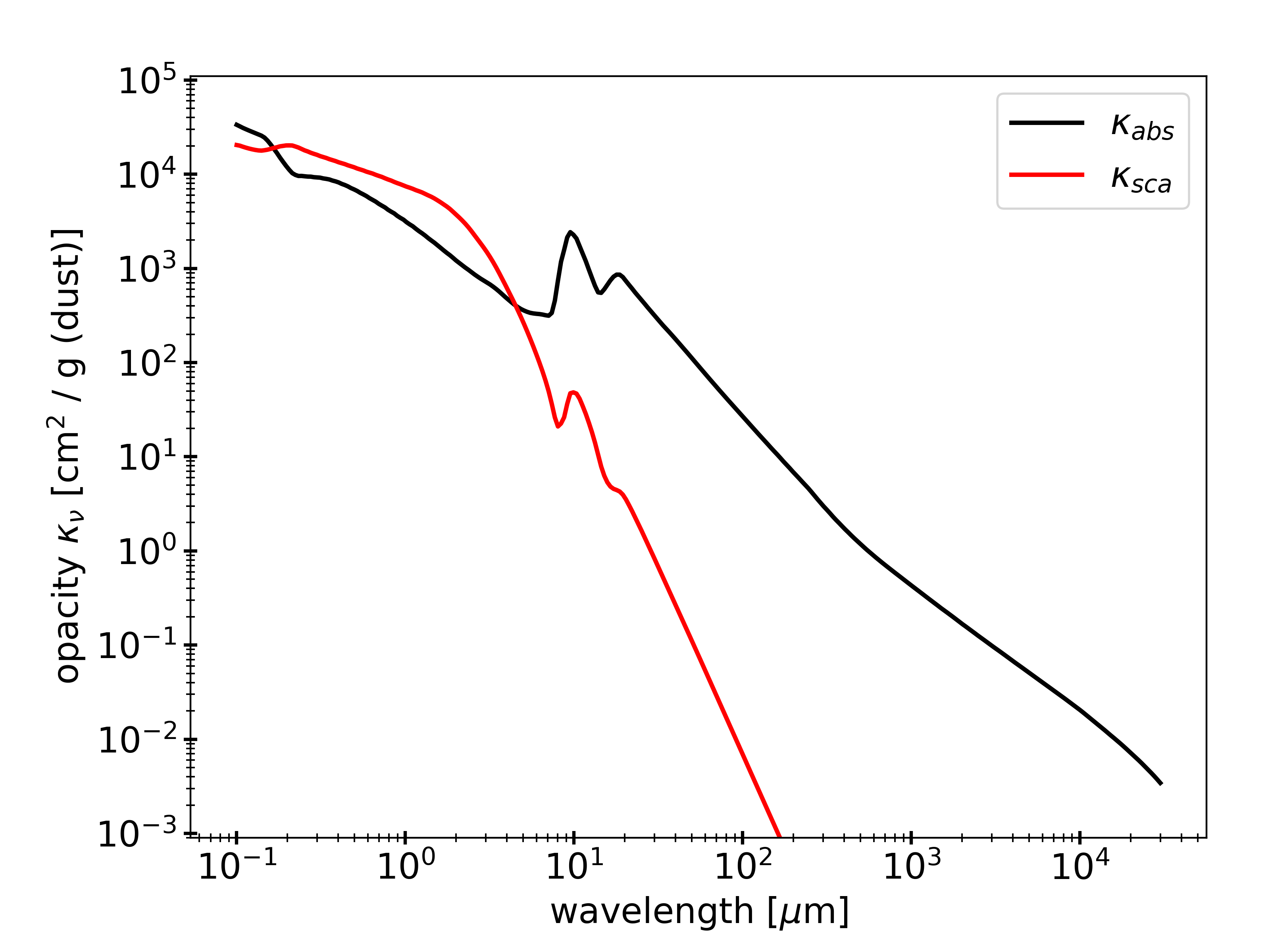}
\end{center}
\caption{Absorption and scattering dust opacities adopted for simulations.}
\label{fig-opacity}
\end{figure}

To model the disk dynamics, we use the \mbox{HURAKAN} RHD code originally described in~\citet{2022ARep...66..800P}. The hydrodynamical solver is based on the classical Godunov method (see, e.g.,~\citet{Kulikovskii:book}) with an exact solution to the Riemann problem. The finite-difference scheme for the spherical coordinate system $(R,\theta)$ is implemented within the formalism described in \citet{Abakumov:2014}, where Godunov-type difference schemes in curvilinear coordinates, their application, and test examples are provided. Radiative transfer is treated with the non-stationary flux-limited-diffusion (FLD) method, with separate treatment for free-streaming radiation from the central source (which we denote as FLD$^{\text{s}}$; see \citet{2024ARep...68.1045P}). In this method, the radiation field is split into high-frequency (stellar + accretion region) and dust thermal components. The heating by the central source radiation is calculated using the frequency-dependent ray-tracing method, while the diffusion approximation with a flux limiter (the FLD approach) and mean opacities are used to treat the propagation of thermal radiation. In our model, we assume that the only source of opacity is dust. The adopted dust opacities correspond to a mixture of spherical silicate and carbonaceous dust particles and are taken from~\cite{2025A&A...700L..24L}, see Fig.~\ref{fig-opacity}.

The simulations are performed on a non-uniform~($R,\theta$) grid with a spatial resolution of 50$\times$150 cells. The computational domain is confined between 2 and 300~au. The inner boundary conditions allow the gas to flow inward. The outer boundary conditions correspond to a rarefied medium that does not influence the disk dynamics over the considered time range.

\section{Results} \label{sec:res}

\begin{figure*}[ht!]
\begin{center}
\includegraphics[width=0.99\linewidth]{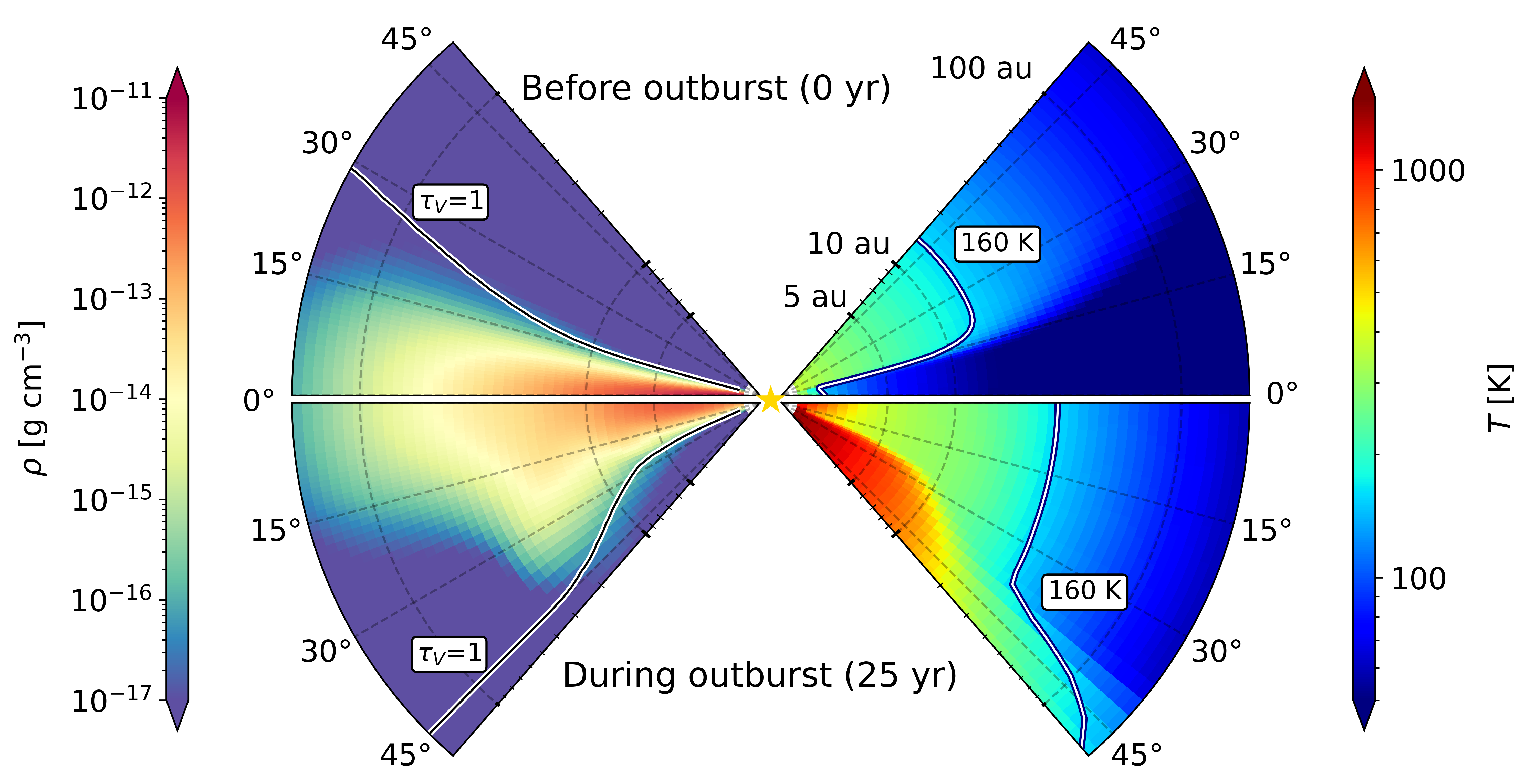}
\end{center}
\caption{Composite view of the protoplanetary disk before and during the outburst. The left and right panels show the density and temperature distributions, respectively, while the upper and lower panels correspond to the pre-outburst and outburst phases. The optical surface to stellar radiation, $\tau_{V}=1$, and the temperature isoline $T=160$~K are also shown.}
\label{chart}
\end{figure*}
{Figure~\ref{chart} provides an overview of the disk response, comparing its density and temperature structure before the outburst and 25~yr after its onset. The distributions are shown in spherical coordinates, with radial distance from the central star plotted on a logarithmic scale. This comparison shows that the outburst substantially alters both the density and temperature structure; we discuss these changes in detail below.
}

\begin{figure*}[ht!]
\begin{center}
    \includegraphics[width=0.99\linewidth]{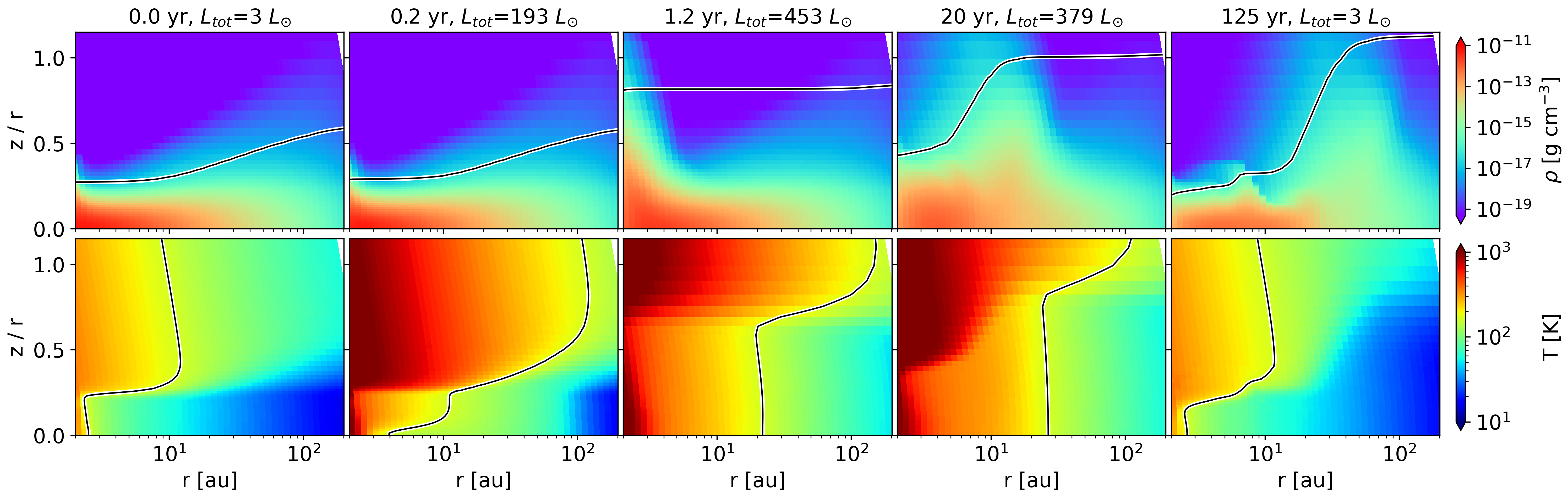}
\end{center}
\caption{Disk density (top row) and temperature (bottom row) distributions at five time moments: before the outburst (left column), 0.2, 1.2, and 20 years after the start of the outburst (middle columns), and 15 years after the end of outburst (right column). The optical surface to stellar radiation, $\tau_{V}=1$, is shown in the top panels. The temperature isoline $T=160$~K corresponding to water snowline is shown in the bottom panels.}
\label{dens-temp}
\end{figure*}

The distributions of density and temperature obtained for different stages of the disk evolution are shown in Fig.~\ref{dens-temp}. The y-axis of these plots, $z/r$,  is chosen  to trace the effects of the shadowing to the stellar radiation: the constant $z/r$ corresponds to the individual ray from the star. The pre-outburst distributions show the classical structure of the passive disk: ultraviolet and optical stellar radiation is absorbed by the disk upper layers, which, in turn, heat the midplane by its own infrared radiation. As a result, the disk is thermally stratified in both the radial and vertical directions. At the inner edge the initial disk is slightly puffed-up due to the direct radiative heating by the central source. 

The temperature distribution at 0.2~yr after the onset of the outburst reveals a significant rise of the temperature both in the envelope and in the interior of the disk. In turn, the density distribution at 0.2~yr is not yet affected since the characteristic dynamical timescale for the inner disk edge is a few times higher. The vertical sound crossing time for a disk in hydrostatic equilibrium is $\sim 1/\Omega_{\rm K}$, which is $0.6$~yr at 2~au distance from $0.6$~M$_\sun$ star.

The key dynamical effect of the outburst seen in the density plots at 1.2 and 20~yr is the development of a density bump travelling outward. The optical surface to stellar radiation ($\tau_{V}=1$, black line in the upper panels of Fig.~\ref{dens-temp}) goes around the bump, so the bump is optically thick and makes the shadow behind it. This is the reason why the temperature behind the bump is lower than the temperature for 0.2~yr at the same location.

Fifteen years after the end of the outburst ($t=125$~yr from its start), the original bump continues to move outwards. However it is now accompanied with a smaller bump(s) generated as disk vertical structure equilibrates to the post-outburst conditions. In the bottom panels of Fig.~\ref{dens-temp}, we also show the 160~K temperature surface corresponding to water sublimation, which is significantly distorted by the outburst. 

\begin{figure*}[ht!]
\begin{center}
    \includegraphics[width=0.95\linewidth]{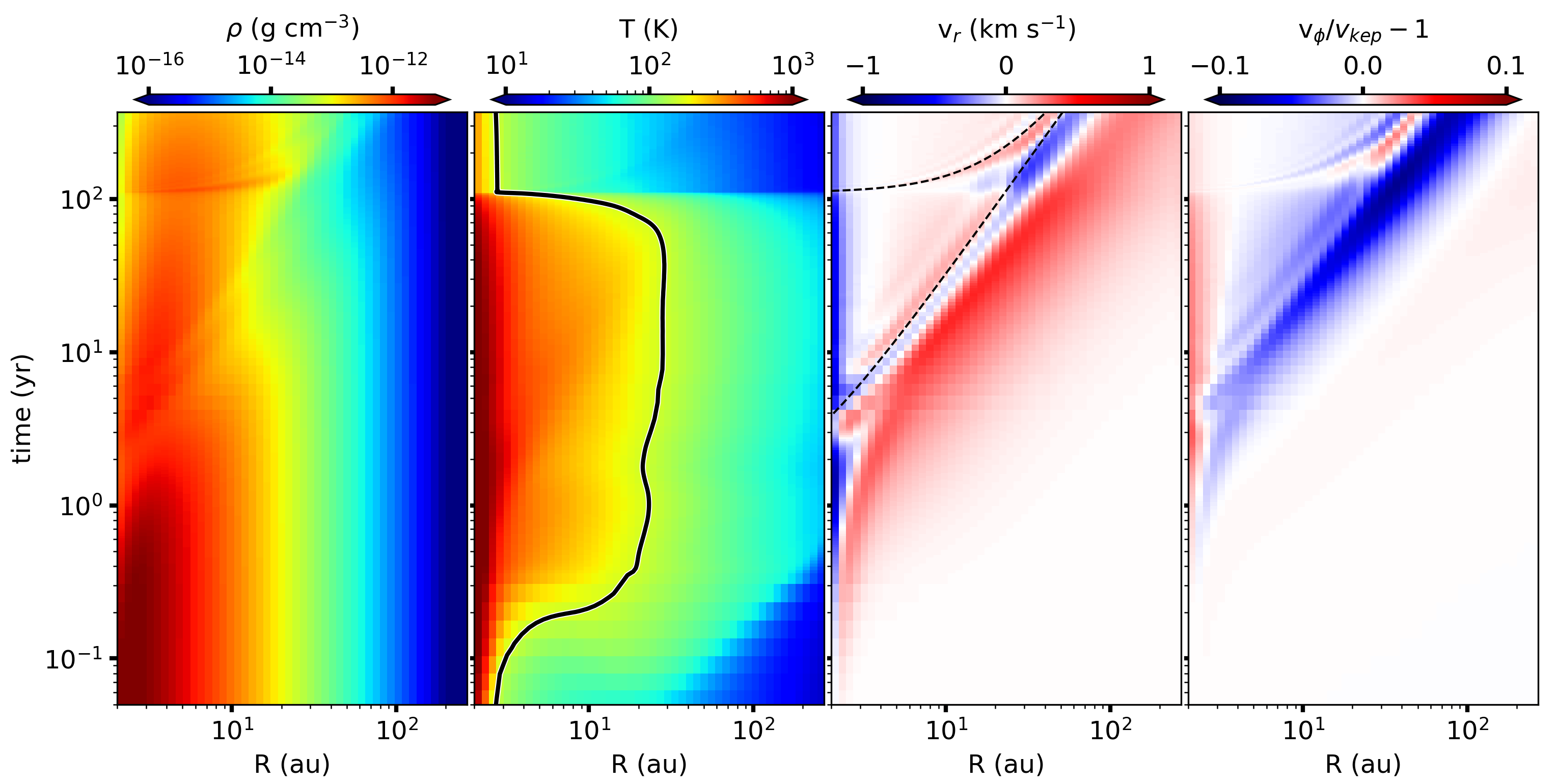}
\end{center}
\caption{Evolution of gas volume density, temperature, radial velocity, and azimuthal velocity in the disk midplane. Black contour corresponds to the water sublimation temperature ($T=160$~K). The power-law fits $r\propto t^{2/3}$ are shown with dashed lines.}
\label{midplane}
\end{figure*}

To show in more detail the disk structure change in time, in Fig.~\ref{midplane} we show the midplane density, temperature, radial velocity, and azimuthal velocity distributions. The heating of the disk due to the outburst during the first 110~years is well seen in the temperature evolution map. The propagation of the initial outburst-driven bump and post-outburst relaxation bump is revealed not only in the density map but even better in the velocity maps. The gas radial velocity associated with moving bump changes the sign from positive to negative and then back to positive. This wave also manifests itself as a decrease in rotation velocity, so the rotation velocity associated with the bump gets substantially sub-keplerian values. The  relative deviation from the keplerian rotation reaches 5\%, which is an order of magnitude larger than the one for a steady state pressure supported disks. 

The radial position of the bump as a function of time is well fitted by the expression $r_{\text{b}} \propto t^{2/3}$ which is plotted by dashed line in Fig.~\ref{midplane} for both the primary outburst and post-outburst bumps. The same power-law, $t_{\text{s}} \propto r^{3/2}$, in fact, is valid for the local sound time, $t_{\text{s}}$, to cross the vertical extent of the disk. Indeed, for the disk in hydrostatic equilibrium, $c_{\text{s}}/v_{\text{K}} \approx h/r$, where $c_{\text{s}}$ is the sound speed, $v_{\text{K}}$ is the keplerian velocity, $h$ is the disk height, so $t_{\text{s}} = h/c_{\text{s}} \approx r/v_{\text{K}} = 1/\Omega_{\rm K}$, where $\Omega_{\rm K}$ is the keplerian frequency. Given this relations, we may conclude that the bump propagation is a wave of disk relaxation to changing accretion luminosity modulated by the local dynamical timescales. 

\begin{figure*}[ht!]
\begin{center}
    \includegraphics[width=0.95\linewidth]{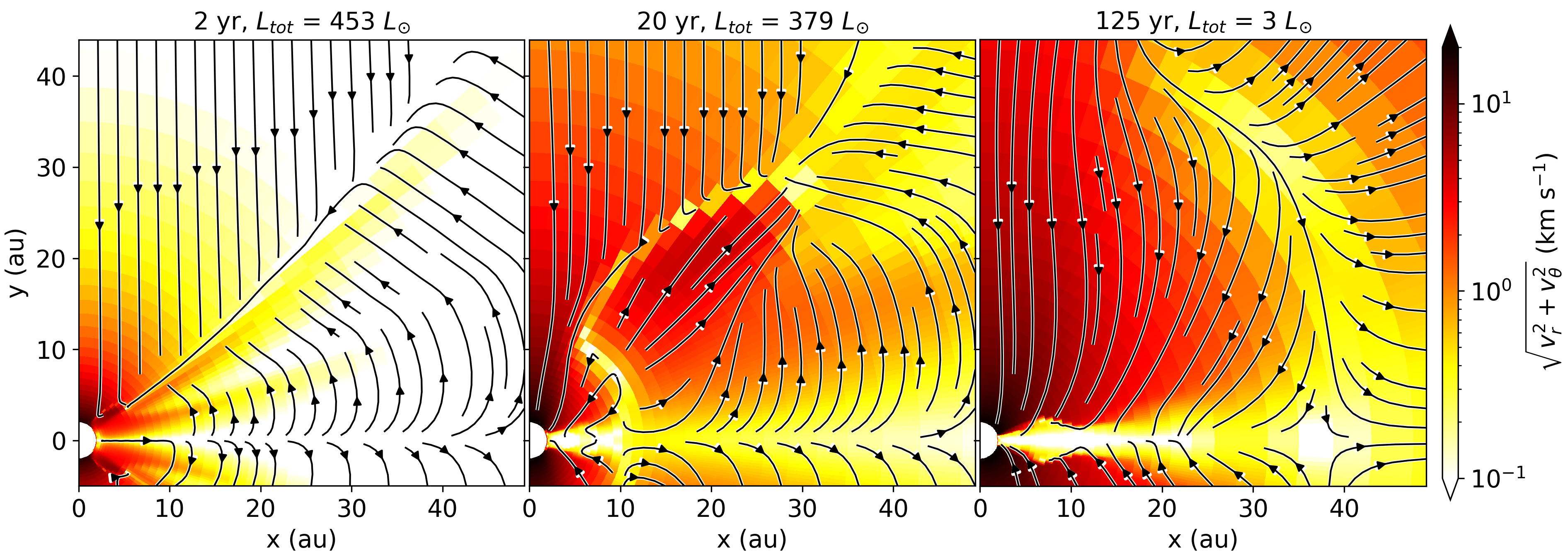}
\end{center}
\caption{Velocity field during (left and middle panels) and after the outburst (right panel). The colour indicates the modulus of the poloidal velocity, while the arrows show its direction.}
\label{vfield}
\end{figure*}

In Fig.~\ref{vfield}, we show the poloidal $\sqrt{v_{r}^2+v_{\theta}^2}$ velocity field distributions within 50~au during and after the outburst. At the beginning of the outburst ($t=2$~yr), the enhanced heating causes the disk to expand in the vertical direction towards the infalling envelope, see the left panel of Fig.~\ref{vfield}. At later outburst stages ($t=20$~yr; the middle panel of Fig.~\ref{vfield}), the disk dynamics involves radial outward flows at disk-envelope interface. Once the outburst ends and the disk cools down ($t=125$~yr, or 15~yr after the end of the outburst; the right panel of~Fig.~\ref{vfield}), the gas loses its vertical pressure support and moves towards the midplane. The primary outburst-driven wave is located at 30~au at this moment and proceeds to travel outwards at later times. Such changes in the velocity structure during and after the outburst are likely to be revealed in the molecular line profiles. Identifying and analysing such molecular line outburst fingerprints requires a separate study. Instead, in this study we focus on observational consequences of the outburst in continuum emission and luminosity curves.

\begin{figure*}[ht!]
\begin{center}
    \includegraphics[width=0.99\linewidth]{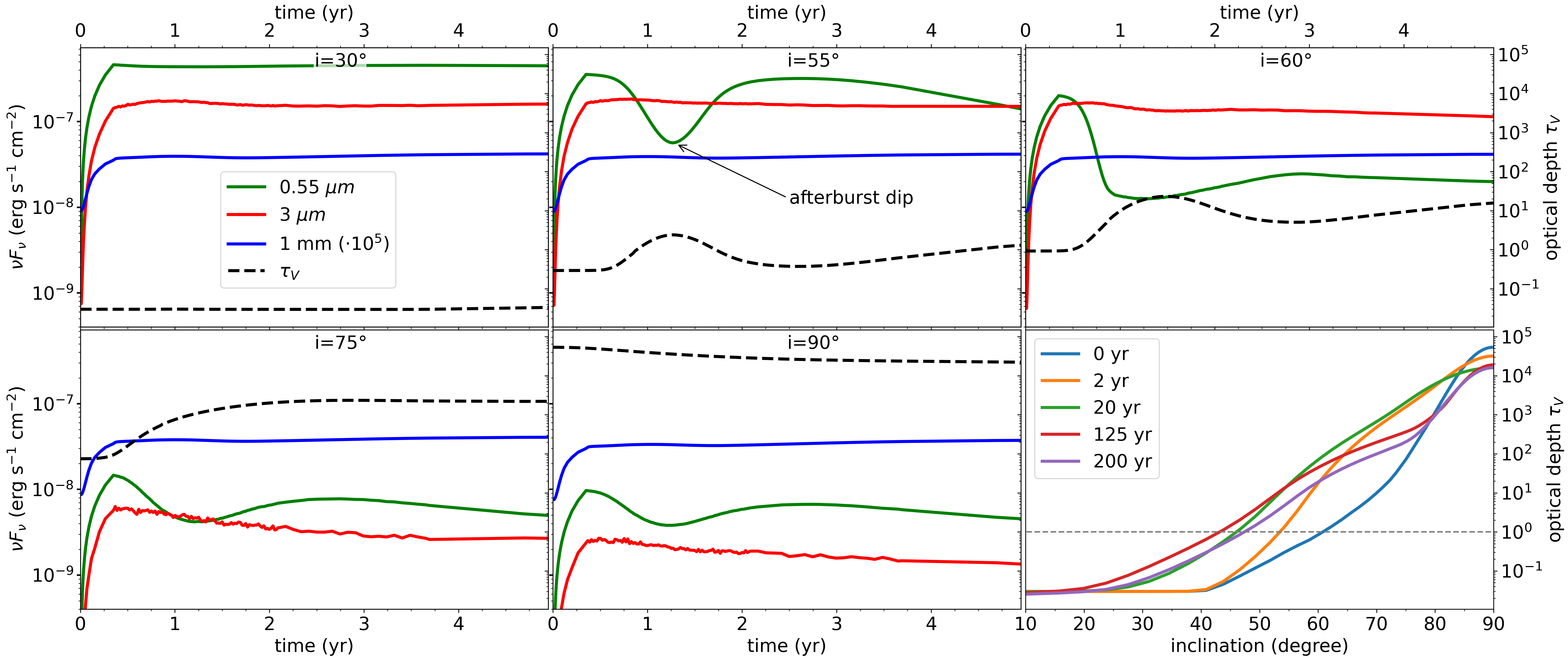}
\end{center}
\caption{Evolution of the fluxes at 0.55~$\mu$m, 3~$\mu$m, and 1~mm for five disk inclinations. Dashed lines indicate the changes in the line-of-sight optical depth towards the star in the V-band, $\tau_V$. Bottom right panel: $\tau_V$ as a function of disk inclination for various stages of the outburst.}
\label{sed-tau}
\end{figure*}

To study the observational signatures of the disk response to the outburst we calculated the variation of total fluxes at 0.55~$\mu$m, 3~$\mu$m, and 1~mm for various disk inclinations, see Fig.~\ref{sed-tau}. The assumed distance to the source is 140 pc. We also calculated how the optical depth in visual band towards the star varies with time and inclination. The fluxes have been computed using the RADMC-3D code\footnote{\url{https://www.ita.uni-heidelberg.de/~dullemond/software/radmc-3d/index.php}}~\citep{2012ascl.soft02015D} at a post-processing step for the density and temperature distributions obtained in the HURAKAN simulations. For the envelope density we assume a floor value of the gas number density of $10^4$~cm$^{-3}$ and the dust-to-gas ratio of 0.01, so that the circumstellar extinction for face-on orientations  $i=0^{\circ}$ is negligible in the considered model setup. Disk expansion in the vertical direction does not affect the visual optical depth towards the star for inclination angles up to $i \approx 40^{\circ}$. As a result, for the low inclined disks the fluxes at all wavelengths quickly (<1 year) become quasi-static and mirror mostly constant accretion luminosity profile during the first 5 years of the outburst (left top panel in Fig.~\ref{sed-tau}). 

For more inclined orientations ($i>40^{\circ}$), the disk expansion in the vertical direction causes substantial interception of the central source emission so that the optical depth at the V band  exceeds unity and significantly varies with time. This results in the apparent flux variation at 0.55~$\mu$m, which is not associated with the intrinsic accretion luminosity variation, see the panels for $i>30^{\circ}$ in Fig.~\ref{sed-tau}. For inclination angles of $i=55^{\circ}, 75^{\circ}$, and $90^{\circ}$, the 0.55~$\mu$m flux exhibits dips by a factor of 2 to 6 around 1.5~yr, corresponding to stellar magnitude variations of $\Delta m\approx0.8-2^m$. At $i=60^{\circ}$, the bump shadowing is even more pronounced and results in a long-lived drop in the apparent optical flux by a factor of 20 ($\Delta m\approx3^m$). Since the optical depths at  3~$\mu$m, and 1~mm are still lower than unity, the temporal evolution of these fluxes is nearly the same as in the case of $i=30^{\circ}$. 

For highly inclined orientation ($i=75^{\circ}$ and larger) the 0.55 and 3~$\mu$m fluxes are  significantly lower due to the high optical depths while at 1~mm the disk is optically thin. The most contribution to the flux at 0.55 and 3~$\mu$m comes from the photon scattering in disk polar regions, which we verified with additional simulations with zero scattering coefficients.

\begin{figure*}[ht!]
\begin{center}
    \includegraphics[width=0.98\linewidth]{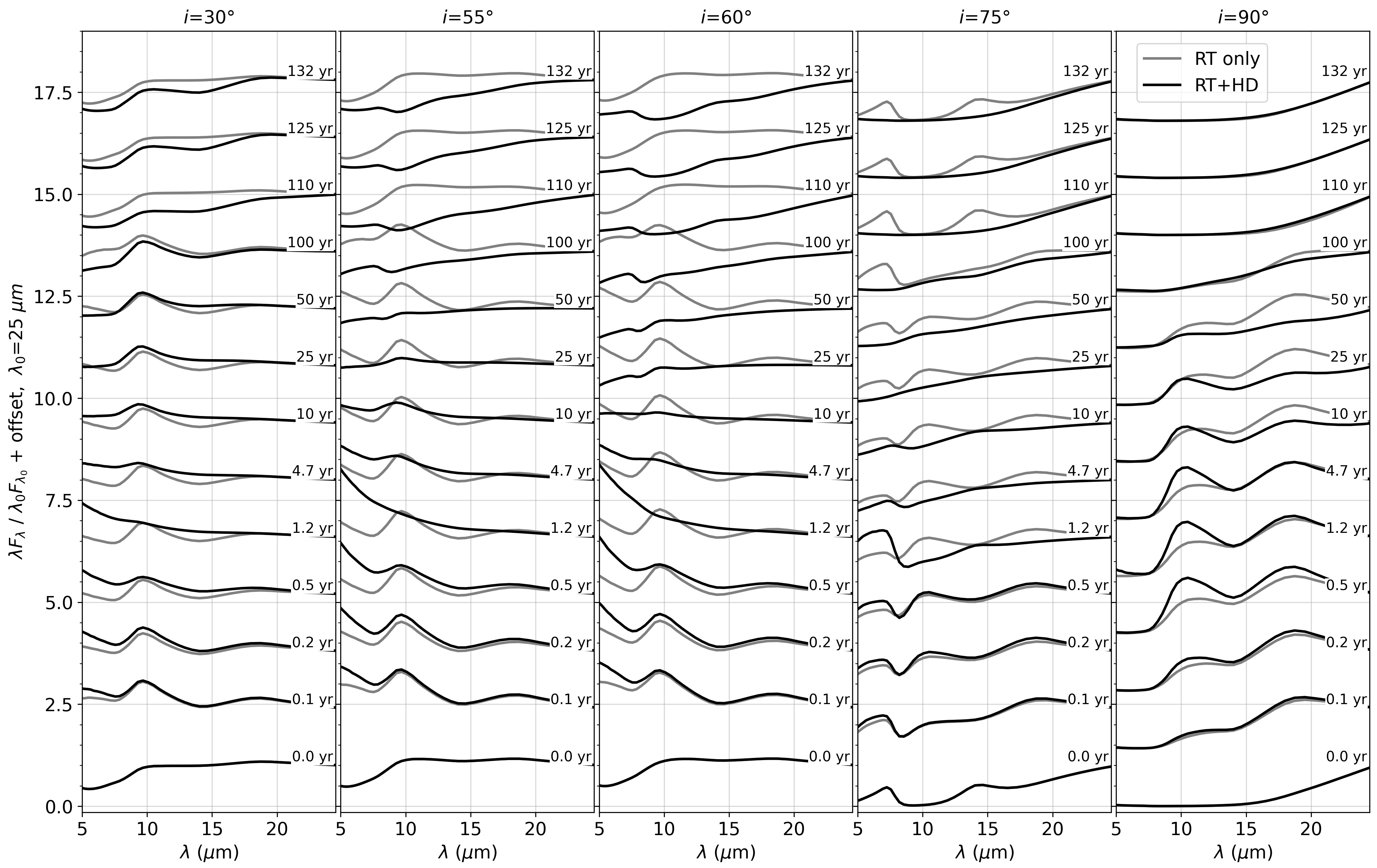}\\
\end{center}
\caption{SED evolution in the $5 - 25\,\mu$m range for the fully dynamical model ('RT+HD')  and the model with density fixed at the pre-outburst values ('RT only'). The SEDs at each time moment are normalized to their $\lambda_0=25\,\mu$m value, and shifted along the y-axis for clarity.}
\label{sed-t}
\end{figure*}

In Fig.~\ref{sed-t}, we show the evolution of the SED between 5 and 25~$\mu$m at various disk inclinations. This wavelength interval includes 10-$\mu$m silicate feature which is an indicator of dust composition and underlying physical conditions. To demonstrate the impact of the disk density changes, we also plot the SED variability for a model without dynamical evolution. This 'RT only' case considers only the time-dependent radiation transfer with the disk density being fixed in the initial quiescent state. As seen from the left column of Fig.~\ref{sed-t} with the low-inclined disk case ($i=30^{\circ}$), both the dynamical and static models show similar SED evolution. The near face-on orientations correspond to the silicate feature in emission, and the difference between 'RT+HD' and 'RT only' models is limited to the transient shadowing of the disk surface in the former model.

At higher inclinations, vertical disk expansion significantly affects the SED, whose evolution differs between the dynamical and static models. In the dynamical model, the 10~$\mu$m feature can appear in emission, absorption, or as a flat continuum, depending on the epoch and inclination. By contrast, in the static model, the feature appears only in emission or as a flat continuum. 
Overall, the SEDs in this wavelength range exhibit a wide variety of shapes, driven by the coupled evolution of the disk's thermal and density structures and modulated by its inclination,
that underscores the need for well-cadenced mid-infrared observations during future outbursts.
A detailed analysis of spectra formation is beyond the scope of this paper and warrants a dedicated study, see e.g.~\citet{2019MNRAS.489.3879B}. As a representative example, Appendix~\ref{AppendixB} examines the spectra from the ``RT-only'' and ``RT+HD'' models 100~yr after the outburst began, at an inclination of $i=75^{\circ}$.

Additional processes not included in our simulations, such as dust evaporation and crystallization, are also likely to influence the SED. Consequently, detailed time-resolved and broadband observations of outbursting sources are essential for robustly constraining disk evolution models.

\begin{figure*}[ht!]
\begin{center}
    \includegraphics[width=0.99\linewidth]{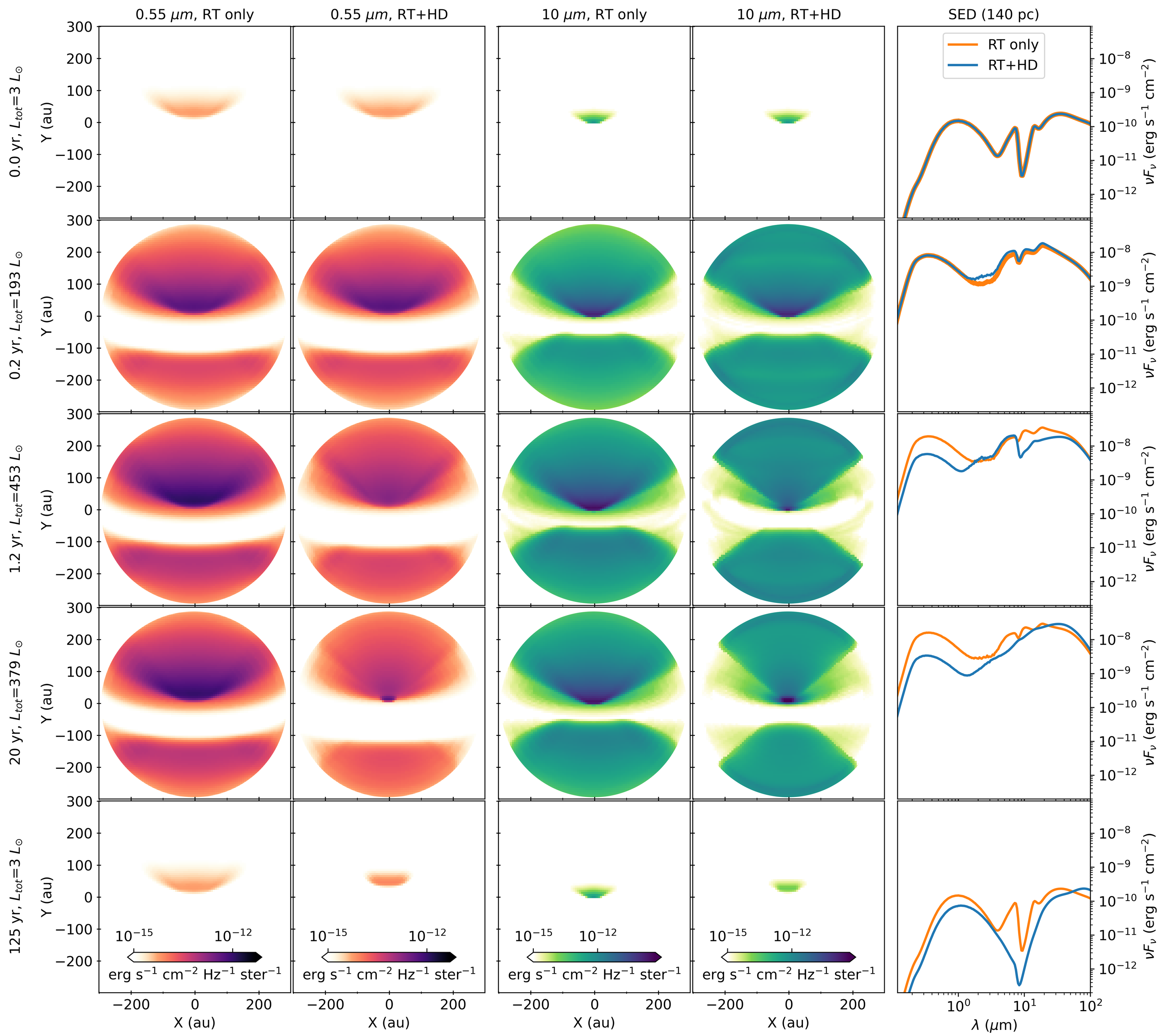}
\end{center}
\caption{Ideal images of the disk at 0.55 and 10~$\mu$m for the static (RT-only) and dynamical (RT+HD) models. The disk inclination is $i=75^{\circ}$. The corresponding time moments are indicated on the left side of each row. The rightmost column shows the SEDs for both models.}
\label{radmc-img-sed}
\end{figure*}

Finally, Fig.~\ref{radmc-img-sed} presents the ideal images of the inclined disk ($i=75^{\circ}$) at 0.55 and 10~$\mu$ m calculated before, during and after the outburst. The images were produced with the RADMC-3D code in the post-processing step, using the temperature and density structures obtained from the HURAKAN simulations. As in the previous figure, we show both the dynamical and static models. The 0.55~$\mu$m emission is dominated by stellar radiation scattered by the disk surface and envelope, whereas the 10~$\mu$m emission is entirely thermal and originates from heated dust. Despite their different physical origins, the morphologies of the 0.55 and 10~$\mu$m images are similar: in both cases, bright bipolar lobes are separated by a dark midplane lane. At 0.55~$\mu$m, the dark lane results from the high optical depth of the midplane to direct radiation from the central source. At 10~$\mu$m, the dark lane is caused by the lower temperatures in the outer midplane regions compared with the envelope, combined with the strong temperature dependence of the Planck function at this wavelength. In the dynamical model, the bipolar lobes have a smaller opening angle and lower brightness than in the static model. This difference is caused by shadowing due to the density bump that develops in the dynamical model. The impact of the emerging bump is clearly reflected in the SED comparison: the 'RT only' model with a fixed density structure exhibits substantially higher fluxes in the visual and near-infrared ranges.

\section{Discussion and Conclusions} \label{sec:concl}
In this work, we investigate the dynamical and thermal response of a protoplanetary disk to an FU Ori-type outburst. We assume that the energy is released in the inner sub-au region, interior to the computational domain spanning $2-300$~au. This setup allows us to study the evolution of the predominantly passive outer disk independently of the specific physical mechanism triggering the outburst.

The disk evolution following the onset of the outburst can be divided into several stages. During the first, relatively short stage, lasting several months and shorter than the characteristic dynamical timescale at the inner disk boundary, the disk is heated by radiation from the central source while its density structure remains nearly unchanged. The second stage is characterized by the dynamical restructuring of the disk in response to the change in its thermal structure. At this stage, an outward-propagating wave is launched from the inner boundary, representing the relaxation of the disk vertical structure toward a new equilibrium corresponding to the enhanced luminosity. The leading front of this relaxation wave, which we refer to as the bump, corresponds to a pronounced local thickening of the disk, with the disk aspect ratio, $h/r$, increasing by a factor of approximately two relative to its pre-outburst value. The position of the bump follows $r_{\rm b}\propto t^{2/3}$, consistent with the radial dependence of the local vertical dynamical timescale. Also, the radial velocities within the relaxation wave reach several km~s$^{-1}$, so the dynamical response involves more than changes in the disk scale height. As the bump propagates outward, it intercepts a fraction of the radiation from the central source, producing a shadowed and consequently cooler region exterior to the bump. The third stage begins with the end of the luminosity outburst, which launches a new relaxation wave propagating outward from the inner boundary. Following its passage, the disk gradually returns to its pre-outburst state on a timescale of $\sim 100$~yr.

The presented model is aimed at investigating the first-order impact of an outburst on the structure of the outer disk and therefore does not account for all potentially relevant physical processes. One of our key assumptions is the separation of the circumstellar disk into inner active and outer passive regions. The influence of the inner active region, where the outburst mechanism operates, is represented solely through the prescribed accretion luminosity profile. In reality, the irradiation of the outer disk may be modulated by the three-dimensional density structure of the actively accreting inner disk, particularly if shadowing by the inner disk wall is important. In Appendix~\ref{AppendixA}, we have shown that attenuation of the passive disk by the inner zones can affect the shape of the bump although it does not affect the overall evolution. 

The physical mechanism driving FU Ori outbursts remains under debate, and their observational manifestations should therefore be studied within the framework of specific outburst-triggering mechanisms \citep{1994ApJ...427..987B, 2001MNRAS.324..705A,2006ApJ...650..956V,2009ApJ...701..620Z, 2026A&A...707A.296C}. Nevertheless, the dynamical response of the outer disk to the increased luminosity may be relatively insensitive to the details of the outburst mechanism. Characterizing this potentially universal response is the main focus of this study.

Several additional assumptions may affect our conclusions regarding the observational manifestations of the outburst. In particular, we do not account for the presence of a circumstellar envelope, disk winds, or bipolar outflows, all of which are likely associated with FU Ori objects. These structures may contribute significantly to both the absorption and scattering of radiation from the central source. The interaction between an outburst-driven disk wind and the infalling envelope, for example, was studied by \citet{2005MNRAS.361..942C}. We also neglect the complex chemical composition of dust grains and their non-uniform spatial distribution within the disk, which may result from radial drift, settling toward the midplane, and thermal processing across snowlines. Consequently, we do not account for changes in dust composition and crystallinity during the outburst, although such changes are expected to occur. As the disk heats up, the regions where icy grain mantles sublimate should expand, potentially producing observable changes in the emergent emission \citep{2025A&A...703L..19W}.

Despite the limitations of the theoretical model, its predictions may provide useful guidance for interpreting observational phenomena. The key observational signature of the relaxation wave identified in our simulations is a substantial, $1 - 2$ order-of-magnitude change in the extinction toward the central source at intermediate disk inclinations of $40^{\circ}-60^{\circ}$. At high inclinations, the outward-propagating bump produces delayed variability in the visible light curve, manifested either as a reduction in the apparent luminosity following an initial maximum persisting for $\sim1$~yr or as a transient dip (see green lines in Fig.~\ref{sed-tau}). For nearly face-on orientations, the line-of-sight extinction variation is minimal, and the outward propagation of the bump has a weaker effect on the apparent light curve.

Although we do not directly compare the presented model with specific FU Ori sources in terms of their observational constraints, some observed objects may exhibit similar behavior. HBC~722 (V2493~Cyg) is a prominent example, where a pronounced dip is observed approximately five months after the outburst maximum \citep{2024CoSka..54b.135S, 2016A&A...596A..52K}. FU~Ori itself showed a $\approx0.5^m$ dip in its $B$-band lightcurve two years after the outburst began \citep{2000ApJ...531.1028K}. PR Ori B provides a more recent example: it showed a $\Delta m_{G}\approx2^m$ minimum between an initial peak and a subsequent brighter maximum \citep{2026arXiv260909722C}. The dynamical response of the disk is also reflected in the variability of the 10~$\mu$m silicate feature. At intermediate disk inclinations, the outward-propagating bump produces substantial variations in the mid-infrared optical depth, causing the silicate feature to evolve from emission to a flat profile or even absorption as the outburst develops. Thus, some FU Ori-type stars exhibiting silicate absorption may be observed during a transient phase and may not necessarily require the presence of a massive circumstellar envelope~\citep{2007ApJ...668..359Q}. In contrast, for face-on and edge-on configurations, the simulations with and without hydrodynamical feedback show qualitatively similar behavior.

\begin{acknowledgments}
We are grateful to the referee for the constructive and stimulating report.
The authors acknowledge the support by the RSF grant 26-12-00349, https://rscf.ru/en/project/26-12-00349/ (Sections 1, 3, 4).
\end{acknowledgments}
\appendix
\section{Bump formation in the inner-disk attenuation model} \label{AppendixA}

The key approximation in our model is to divide the protoplanetary disk at the inner computational boundary, $R_{\rm min}$, into an inner active disk and an outer passive disk. We assume that instabilities in the inner disk can enhance accretion onto the star and trigger the outburst, while the outer disk does not undergo strong instabilities and responds only to the outburst radiation. We therefore exclude the inner active disk from the computational domain and do not model its physics. Instead, we represent the stellar and inner disk emission with two black-body components with prescribed parameters. In the base model described in Section~\ref{sec:res}, radiation from the central source directly heats the inner edge of the outer disk, potentially affecting its dynamics and response to the outburst. Because the actual disk is continuous, material in the inner disk may attenuate this radiation before it reaches the modeled outer disk. The base model may therefore overestimate the heating at the outer disk inner edge.

To test whether this artificial division affects the formation and propagation of the bump, we also consider an inner-disk attenuation model that accounts for absorbing material inside $R_{\rm min}$. In this model, the central source flux is reduced by a factor $\exp{(-\tau_{\nu})}$, where
$\tau_{\nu}=\kappa_{\nu}\rho_{\rm in}(R_{\rm min} - R_0)$ represents the additional optical depth between the presumed inner disk edge, $R_0$, and the computational grid boundary, $R_{\rm min}$. We adopt $R_0=0.5R_{\rm min}$ as a rough estimate (see \citet{2017ApJ...850..131F} for a similar approach). For each radial ray at angle $\theta$, we take the density $\rho_{\rm in}$ from the second radial cell (taking the first radial cell leads to unstable initial density-temperature hydrostatic iterations). We then extrapolate the disk density inward to estimate attenuation by the inner disk, which is not included in the computational domain.

We also consider two limiting boundary conditions for the disk thermal radiation: (A1) free escape through the inner edge, i.e., $F=-cE_r$, where $F$ is the radial component of the radiation flux and $E_r$ is the radiation energy density; and (A2) full reflection, i.e., $F=0$. The actual condition is expected to lie between these limits.

\begin{figure*}[ht!]
\begin{center}
    \includegraphics[width=0.95\linewidth]{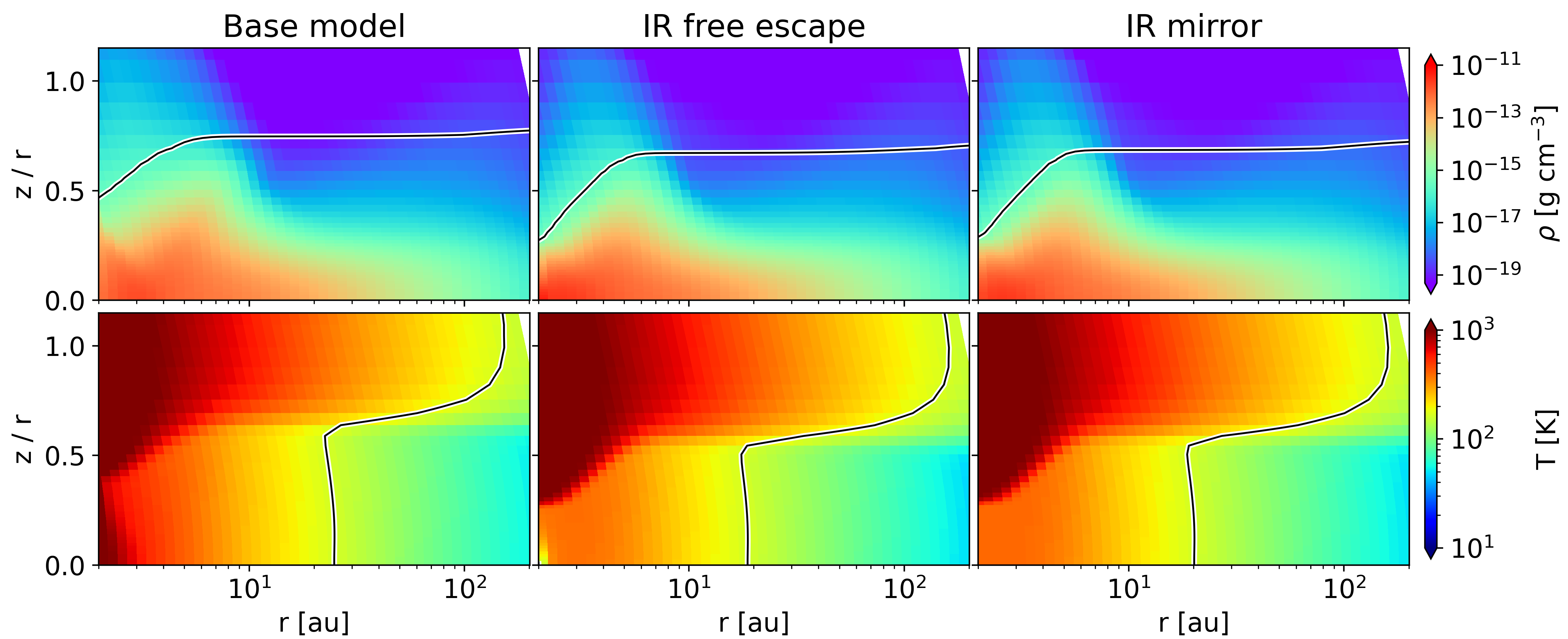}
\end{center}
\caption{Density and temperature distributions 10~yr after the start of the outburst for the base model (left), the attenuation model with a free escape boundary for infrared radiation (middle), and the attenuation model with a reflecting boundary for infrared radiation (right).}
\label{fig-attenuated}
\end{figure*}

\begin{figure*}[ht!]
\begin{center}
    \includegraphics[width=\linewidth]{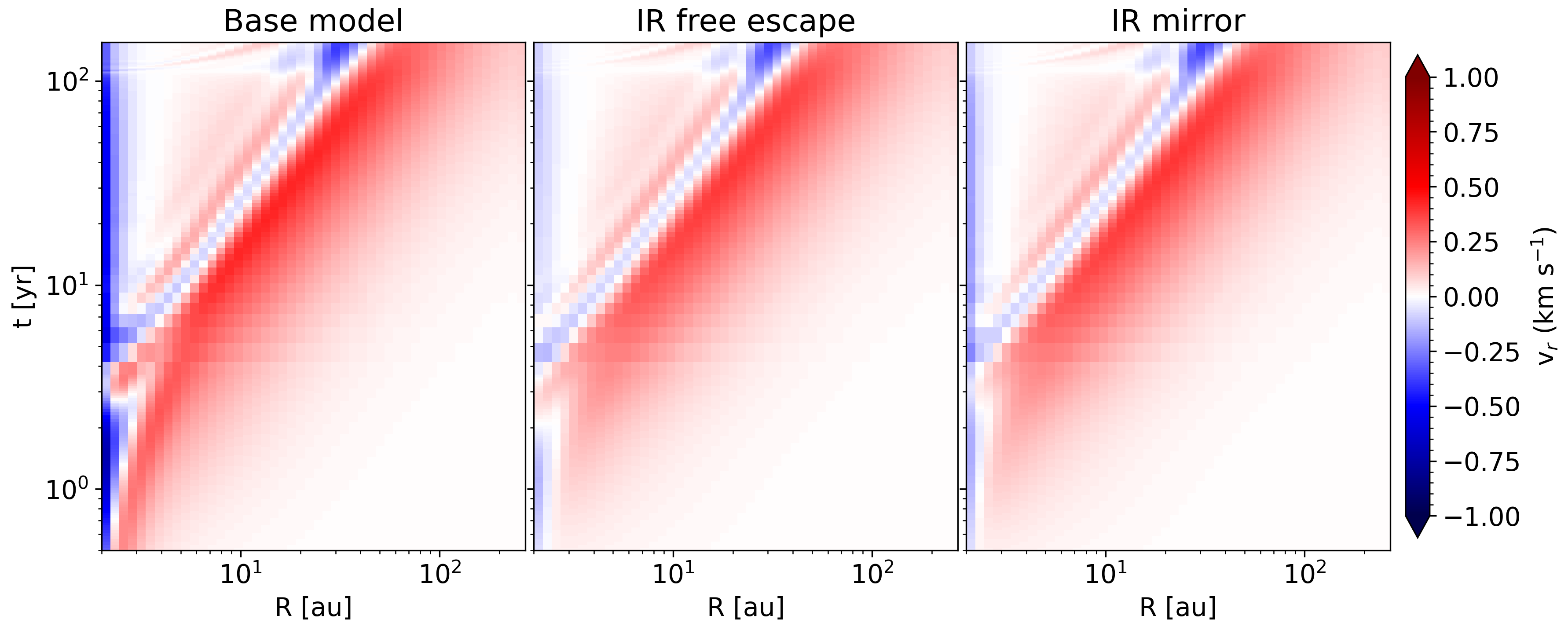}
\end{center}
\caption{Evolution of the radial velocity in the base model (left) and in the inner-disk attenuation models with free escape (middle) and reflecting (right) boundary conditions for thermal infrared radiation.}
\label{fig-attenuated-vr}
\end{figure*}

Figure~\ref{fig-attenuated} shows the two-dimensional density and temperature distributions 10~yr after the onset of the outburst for the base model and the two attenuation models. Near the inner disk edge, the temperature peaks in the base model, reaches a minimum in model A1, and forms a plateau in model A2. As expected, temperatures near the edge are lower in both attenuation models than in the base model. The choice of inner boundary condition modestly affects the shape of the bump but does not alter the disk global dynamics. This is reflected in Fig.~\ref{fig-attenuated-vr}, where the midplane radial velocity evolves nearly identically in all three models.

\section{Analysis of the spectra near 10~micron feature} \label{AppendixB}

To investigate the differences between the ``RT-only'' and ``RT+HD'' SEDs near the 10~$\mu$m feature shown in Fig.~\ref{sed-t}, we focus on the spectra at $t=100$~yr and an inclination of $i=75^{\circ}$. In the ``RT-only'' model, the spectrum has a local maximum at 7.5~$\mu$m, whereas in the ``RT+HD'' model, the SED is nearly flat from 7.5 to 9.5~$\mu$m.

\begin{figure*}[ht!]
\begin{center}
    \includegraphics[width=0.95\linewidth]{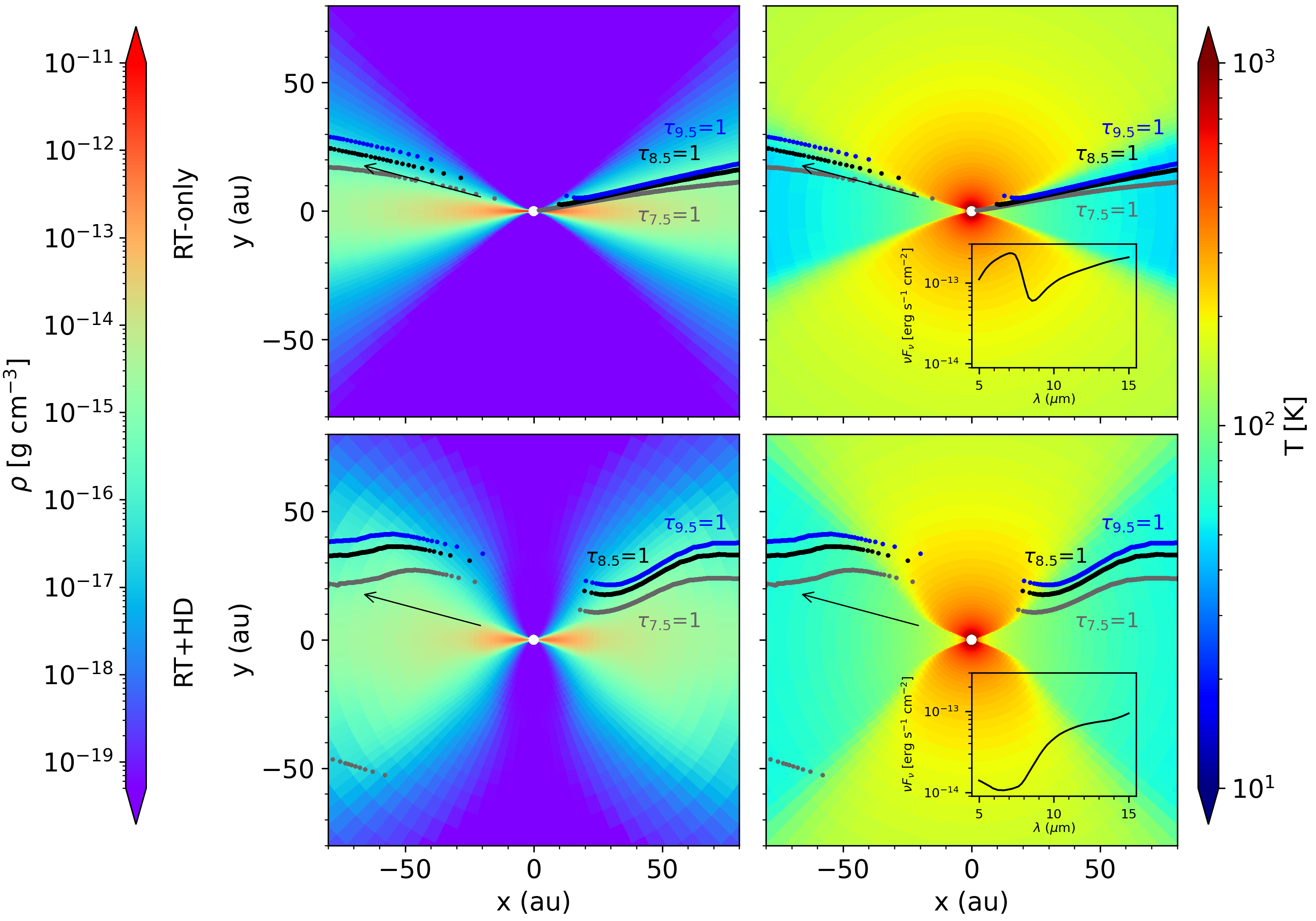}
\end{center}
\caption{Density (left column) and temperature (right column) distributions 100~yr after outburst onset for the ``RT-only'' (top row) and ``RT+HD'' (bottom row) models. Gray, black and blue dots indicate the $\tau=1$ surfaces at 7.5, 8.5, and 9.5~$\mu$m respectively. The viewing angle is $75^{\circ}$ from the polar axis, and the arrow marks the observer's direction. The SEDs from 5 to 15~$\mu$m are overlaid on the right-hand panels.}
\label{fig-AppendixB}
\end{figure*}

 In Fig.~\ref{fig-AppendixB}, we plot the density and temperature distributions together with the $\tau=1$ surfaces for 7.5, 8.5, and 9.5~$\mu$m. The optical depth  is calculated from the observer. At these wavelengths, the absorption opacity exceeds the scattering opacity, so thermal processes dominate emission. Using these plots we rely on the useful approximation, $I_{\nu} \approx B_{\nu}(T_{\rm surf})$, where $I_{\nu}$ is the emergent intensity, $B_{\nu}$ is the Planck function, $T_{\rm surf}$ is the temperature on corresponding $\tau_{\nu}=1$ surface. The optical depth surfaces are calculated using RADMC-3D tools.

The unit optical depth surfaces are vertically ordered by wavelength: the $\tau_{9.5}=1$ surface lies above the $\tau_{8.5}=1$ surface, which in turn lies above the $\tau_{7.5}=1$ surface. This ordering is consistent with the increasing opacity from 7.5 to 9.5~$\mu$m.

In the ``RT-only'' model, the unit optical depth surface approaches the disk inner edge on its far side, whereas the $\tau_{8.5}=1$ and $\tau_{9.5}=1$ surfaces reach radial distances of approximately 15 and 20~au, respectively. As a result, $7.5$ $\mu$m emission comes from the hot inner disk edge providing the local peak in the SED. 

In the ``RT+HD'' model, the unit optical depth surfaces lie much higher above the midplane than in the ``RT-only'' model. The density bump obscures the innermost disk from the observer, and all three surfaces begin at approximately 25~au. As a result, the SED has a lower flux throughout the $5-15~\mu$m range. Because the surfaces pass through regions with similar temperatures, the SED varies monotonically across this wavelength range.

\begin{figure*}[ht!]
\begin{center}
    \includegraphics[width=0.95\linewidth]{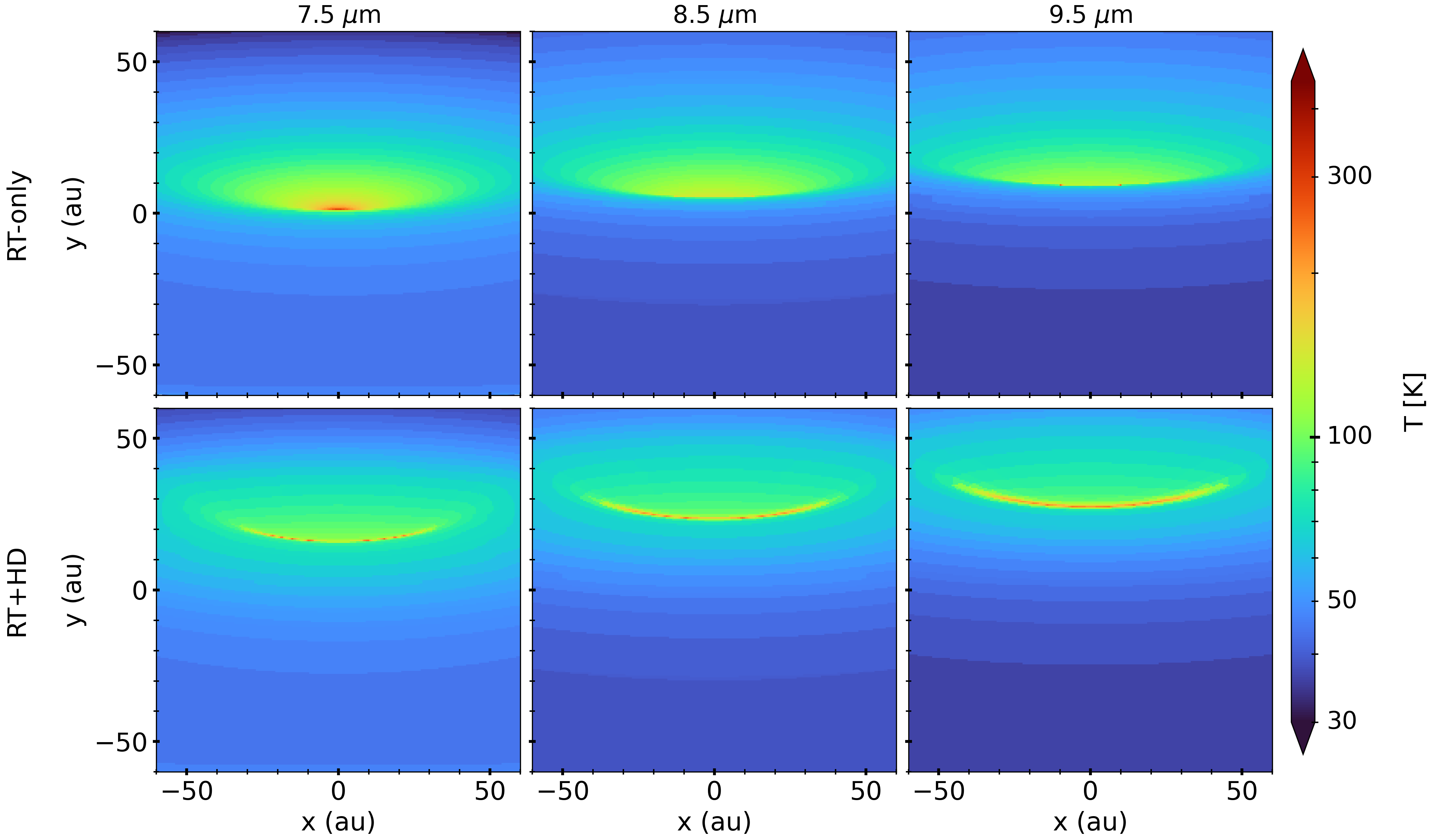}
\end{center}
\caption{The temperature distributions on the unit optical depth surfaces, projected onto the image plane.}
\label{fig-AppendixB2}
\end{figure*}

As a complementary check, Fig.~\ref{fig-AppendixB2} shows the temperature distributions on the unit optical depth surfaces, projected onto the image plane. These emitting surfaces are color-coded by their local temperature, $T_{\rm surf}$. Under the approximation $I_{\nu} \approx B_{\nu}(T_{\rm surf})$, warmer (redder) regions contribute more strongly to the emergent flux at a given frequency.

In the ``RT-only'' model, the 7.5~$\mu$m map shows a hot inner region that contributes substantially to the total flux. The hot inner region is absent on the 8.5 and 9.5~$\mu$m maps, explaining the local maximum in the SED at 7.5~$\mu$m.

By contrast, the temperature distributions in the ``RT+HD'' model have similar, arc-like morphology at all three wavelengths. The bright (red) narrow arcs in these maps appear since the unit optical depth curve passes through the hot polar regions, see few points in the left side of the disk in Fig.~\ref{fig-AppendixB}. Because the warm regions have similar temperatures and areas, the SED remains nearly flat across $7.5-9.5\,\mu$m. The emitting layers in the ``RT+HD'' model also lie higher than those in the ``RT-only'' model.

\bibliography{bibliography}{}
\bibliographystyle{aasjournalv7}
\end{document}